\documentclass[aps,prd,preprint,superscriptaddress,nofootinbib,showpacs]{revtex4-2} 
\usepackage{graphicx,amsmath,amsfonts,amssymb,revsymb,dcolumn,relsize,hyperref}
\hypersetup{colorlinks=true,citecolor=blue,urlcolor=blue,linkcolor=blue}

\newcommand{\be}{\begin{equation}}
\newcommand{\ee}{\end{equation}}
\newcommand{\bea}{\begin{eqnarray}}
\newcommand{\eea}{\end{eqnarray}}

\begin{document}

\title{G\"odel-type spacetimes in $f(Q)$ gravity}
\author{A. M. \surname{Silva}} \email{alisson.matheus@unesp.br}
\affiliation{Instituto de F\'{\i}sica Te\'{o}rica, UNESP-Universidade Estadual Paulista, R. Dr. Bento T. Ferraz 271, Bl. II,
\\
01140-070 S\~ao Paulo -- SP, Brazil }
\author{M. J. \surname{Rebou\c{c}as}} \email{reboucas.marcelo@gmail.com}
\affiliation{Centro Brasileiro de Pesquisas F\'{\i}sicas,
Rua Dr.\ Xavier Sigaud 150 \\
22290-180 Rio de Janeiro -- RJ, Brazil}

\author{N. A. \surname{Lemos}} \email{nivaldolemos@id.uff.br}
\affiliation{Instituto de F\'{\i}sica, Universidade Federal Fluminense,
Av. Litor\^anea, S/N \\
24210-340 Niter\'oi -- RJ, Brazil }




\renewcommand{\baselinestretch}{0.96}

\date{\today}

\begin{abstract}

There is ongoing interest in the nonmetricity formulation of gravity. The nonlinear extension of the theory, called $f(Q)$ gravity, has recently been proposed and offers a promising avenue for addressing some of the long-standing challenges in cosmology and fundamental physics. A number of solutions have already been found that have confirmed the usefulness of the theory for astrophysics and cosmology.
Motivated by earlier work on G\"odel-type spacetimes
we investigate whether $f(Q)$ gravity admits  such cosmological solutions, which entail causality violation. In the coincident gauge, which is shown to be a legitimate gauge because the  connection  field equations are satisfied, we find that G\"odel-type metrics are allowed by this  modified gravity theory for any function $f(Q)$. Noncausal and causal solutions exist depending on the matter content. For a perfect fluid,   out of all G\"odel-type metrics only G\"odel's original model is allowed in $f(Q)$ gravity, implying violation of causality. The requirement that energy conditions be obeyed by the fluid leads to mild restrictions on $f(Q)$. For a massless scalar field in the presence of a cosmological constant,   there are causal solutions  for any reasonable function $f(Q)$.
 \end{abstract}


\maketitle
\newpage

\section{Introduction}

General relativity is a well-established  theory that has successfully passed  many stringent  experimental tests \cite{Will}.  However, the application of Einstein’s theory of gravity to cosmology requires that  the universe contain  dark matter and be uniformly filled with dark energy,  which together comprise around 95\% of the universe's material content but whose  physical nature  remains mysterious. Inflation is also an external ingredient that is added to the very early stages of the universe's evolution in order to solve certain fine tuning problems \cite{Maroto}. 

It is reasonable to assume that, rather than adding exotic components to the matter content, the challenges of the standard cosmological model can be met through some modification of general relativity that still withstands all solar system  and astrophysical tests.  The fact that the canonical quantization of general relativity seems to give rise to a nonrenormalizable quantum field theory \cite{Hooft,Goroff} further stimulates the search for a modified theory that might be more amenable to quantization.
The teleparallel equivalent of general relativity is a valid candidate to providing a more favourable framework to canonical quantum gravity \cite{Ulhoa,Kanatchikov,Aldrovandi-Vu,Aldrovandi-Gui},  one of the important reasons being that in   teleparallel gravity the energy-momentum of the gravitational field is described by a true tensor \cite{nester,Aldrovandi}. Thus, modified theories of gravity offer a viable avenue to bridge the gap between macroscopic and microscopic gravity, hopefully providing a more comprehensive and unified framework for understanding both the early and  late stages of the universe's evolution \cite{modified1,modified2,modified3}. 

Among other possibilities,  general relativity  can be equivalently formulated in terms of the nonmetricity tensor \cite{nester}, which measures the departure from zero of the covariant derivative of the metric tensor (see \cite{Lavinia-review} for an extensive review). In this approach, the Riemann tensor and the torsion tensor  are both zero but the connection is not necessarily the Levi-Civita connection (Christoffel symbols). The gravitational effects are ascribed to the nonmetricity tensor. In this theory,  the curvature scalar $R$ is shown to differ only by a total derivative from a scalar $Q$ constructed from the nonmetricity tensor. This means that the Einstein-Hilbert action with $R$ replaced by $Q$ leads to the standard Einstein equations of general relativity. However, if $R$ is replaced by $f(Q)$ the modified Einstein-Hilbert action leads to new field equations that are not equivalent to Einstein's equations as long as the second derivative of $f$ does not vanish. Beside the metric field equations there are separate connection field equations that  also must be satisfied. This theory, known as $f(Q)$ gravity, has been under intense investigation during the past few years both at the classical level (see \cite{Lavinia-review} and references therein) and in quantum cosmology \cite{Christodoulakis,Capozziello}.
Novel solutions, differing from those of general relativity, have been obtained in $f(Q)$ gravity both in cosmology \cite{cosmology1,cosmology2} and in the spherically symmetric case \cite{Lin-2021,Zhao-2022,Lavinia-BlackHole,Calza}.  If $f(Q)= AQ + B$, with $A$ and $B$ constants, the theory is equivalent to general relativity with a cosmological constant.  

Since $f(Q)$ gravity assumes that the Riemann tensor vanishes, it is possible to choose the connection in such a way that all of its components are zero and the covariant derivative reduces to the ordinary partial derivative \cite{lavinia1}. This choice   defines the  so-called coincident gauge. In fact there is a set of privileged coordinate systems --- each two of them related by an affine coordinate transformation --- in which the connection is zero. 
However, it is not clear if a certain coordinate system chosen to express the metric is one of the privileged coordinate systems in which the connection vanishes.  This raises the question of compatibility between the coincident gauge and a particular coordinate system in which one is working. The compatibility criterion we adopt is straightforward: for the coincident gauge to be compatible with the chosen coordinate system the trivial connection $\Gamma^{\lambda}_{\,\,\mu \nu} =0$ must obey the connection field equations.

In 1949 Kurt G\"odel found a cosmological solution to Einstein's equations with surprising features.   G\"odel's solution not only describes a rotating universe but also leads to causality violation in the form of closed timelike curves \cite{Godel}. Ever since G\"odel's discovery, there have been quite a few studies of  G\"odel-type spacetimes \cite{Raychaudhuri,Reboucas-tese,Reb_Tiomno,Teixeira,Smalley,Reboucas-Note,RebAman,Reboucas-Riem,Reboucas-Cartan,Reboucas-Riem-Cartan,Kanti,Reboucas-Ind-Matter,Barrow, Reboucas-Chron,Agudelo,daSilva,Boskoff}. In particular, G\"odel-type solutions have been investigated in nonlinear extensions of general relativity  such as $f(R)$ gravity \cite{Reboucas-f(R),marcelo-fr,Santos-Reb}, $f(T)$ gravity where $T$ is the torsion scalar \cite{marcelo-ft},  $f(R,T^2)$ gravity in which the action depends on the energy-momentum tensor squared \cite{Canuto-Santos}, and $f(R,Q,P)$ gravity where $Q$ and $P$ are respectively quadratic in the Ricci and Riemann tensors \cite{Nascimento}.

 In this paper we look for possible G\"odel-type solutions and the associated causality violation in  $f(Q)$ gravity. The metrics of interest are cylindrically symmetric and we investigate if they are consistent with nontrivial $f(Q)$ gravity. In the coincident gauge, which is proved to be a legitimate choice because the connection field equations are satisfied, we find that    $f(Q)$ gravity does admit G\"odel-type metrics for any function $f(Q)$. If the matter content consists of a perfect fluid, out of all G\"odel-type metrics only G\"odel's original model is compatible with $f(Q)$ gravity. This implies that violation of causality in the form of closed timelike curves is a feature common to general relativity and  $f(Q)$ gravity. We also find that only mild restrictions on the function $f(Q)$ arise from the imposition of energy conditions on the perfect fluid. If the matter content is described by a massless scalar field and there is a cosmological constant, there exist causal solutions for any reasonable function $f(Q)$.

In Section \ref{f(Q)gravity} we give a brief summary of $f(Q)$ theory. In Section \ref{Godeltype} we present the G\"odel-type metrics. In Section \ref{Godeltype-f(Q)} we set up the field equations of $f(Q)$ gravity in the coincident gauge, which is shown to be consistent with the connection field equations.  In Section \ref{Perfectfluid} we show that for a perfect fluid
 only G\"odel's original model is allowed in $f(Q)$ gravity. Therefore,  the theory entails  causality violation in the form of closed timelike curves. In Section \ref{Energycond} it is found
 that the imposition of energy conditions on the fluid leads just to weak restrictions on the function $f(Q)$. In Section   \ref{Scalar field} it is shown that for a massless scalar field in the presence of a cosmological constant there are causal solutions for any sensible function $f(Q)$.  Section   \ref{Conclusions} is devoted to general remarks on our main findings. 
The results of lengthy computations are consigned to the Appendix so as not to clutter up the main text.
 

\section{A brief overview of $f(Q)$ gravity}\label{f(Q)gravity}

One approach to reasonable modifications of Einstein's theory of gravity involves considering different geometrical objects in the gravitational action, such as the torsion scalar $T$ in the theory of teleparallelism  or the nonmetricity scalar $Q$ in the theory of coincident general relativity \cite{nester,lavinia1}. A unifying framework to describe these modifications of general relativity was introduced in \cite{trinityl}, where   a triangle  was considered with each vertex representing one of the possible fundamental geometrical objects in terms of which the theory can be formulated. This so-called geometric trinity summarizes three equivalent ways to describe general relativity.

Of particular interest to the present article is the nonlinear extension of the theory of coincident general relativity
 \cite{nester,lavinia1},   also known as symmetric teleparallel general relativity (STGR). Its fundamental object is the nonmetricity tensor $Q_{\alpha \mu \nu}$ which is defined by
\begin{equation}
\label{Nonmetricity}
Q_{\alpha \mu \nu}=\nabla_\alpha g_{\mu\nu} = \partial_{\alpha}g_{\mu\nu} - \Gamma^{\lambda}_{\,\,\, \alpha\mu} g_{\lambda\nu}- \Gamma^{\lambda}_{\,\,\, \alpha\nu} g_{\mu\lambda}
\end{equation}
and measures the variation of the norm of a vector field under parallel transport specified by the symmetric connection $\Gamma^{\lambda}_{\,\,\, \mu\nu}$. The assumption that $Q_{\alpha\mu\nu}$ does not vanish makes room for a different geometric interpretation  of gravitational interactions, focusing on the distortion of lengths rather than angles \cite{Lavinia-review}. Since it is also postulated that the Riemann tensor as well as torsion are zero, the connection is symmetric and the gravitational effects are encoded in the nonmetricity tensor.



\ 

Nonlinear extensions of the previously mentioned theories have long been investigated, first in the form of $f(R)$ gravity,  which takes into account nonlinear terms in the Ricci scalar $R$, and next in the form of $f(T)$ gravity which considers nonlinear terms in the torsion scalar $T$. The recently proposed theory of $f(Q)$ gravity containing nonlinear terms in the nonmetricity has been drawing a lot of attention.
Similarly to what is done in $f(R)$ and $f(T)$ theories, upon introducing a general infinitely differentiable function $f(Q)$ of the nonmetricity  scalar one is led  to an action of the form \cite{lavinia1} 
\begin{equation}
\label{action}
S = \frac{1}{2}\int d^4 x \sqrt{-g}f(Q) + S_m,  
\end{equation}
where $8\pi G/c^4 =1$ and $S_m$ is the matter action.
Varying this action with respect to the metric leads to the metric field equations \cite{cosmology1}
\begin{equation}
\label{f(Q)-equations}
 \frac{2}{\sqrt{-g}}\nabla_\alpha\bigl(\sqrt{-g}f_QP^{\alpha}_{\,\,\,\,\mu \nu}\bigr)- \frac{1}{2}fg_{\mu \nu}+f_Q\bigl(P_{\nu \alpha \beta}Q_{\mu}^{\,\,\,\alpha \beta}-2P_{\alpha \beta\mu}Q^{\alpha \beta}_{\,\,\,\,\,\,\,\, \nu}\bigr) = T_{\mu \nu} 
\end{equation}
where the subscript $Q$ means derivative: $f_Q(Q) = f^{\prime}(Q)$. 
Variation of the action \eqref{action} with respect to the connection yields the connection field equations \cite{cosmology1}
\begin{equation}
\label{connection-equations}
 \nabla_{\mu}\nabla_{\nu} \bigl( \sqrt{-g}f_Q P^{\mu\nu}_{\quad \alpha}\bigr) =0,
\end{equation}
which are dynamical equations for the connection.
In the above equations $P^{\alpha\mu \nu}$ is the nonmetricity conjugate tensor defined as
\begin{equation}
\label{P-alpha-mu-nu}
    P^{\alpha\mu \nu}=-\frac{1}{4}Q^{\alpha\mu \nu}+\frac{1}{2}Q^{(\mu\nu)\alpha}+\frac{1}{4}(Q^{\alpha}-{\tilde Q}^{\alpha})g^{\mu \nu}-\frac{1}{4}g^{\alpha (\mu}Q^{\nu)}
\end{equation}
where the parentheses surrounding the indices denote symmetrization, as for example in $A_{(\mu \nu )}=(A_{\mu\nu}+A_{\nu\mu})/2$.
In the preceding equation there also appear the independent traces of the nonmetricity tensor given by
\begin{equation}
\label{Q-tildeQ}
    Q_{\alpha} = g^{\sigma \lambda}Q_{\alpha \sigma \lambda}, \qquad {\tilde Q}_{\alpha} = g^{\sigma \lambda}Q_{\sigma \alpha \lambda}.
\end{equation}
Finally, the   nonmetricity scalar $Q$ is defined by
\begin{equation}
\label{nonmetricity-scalar-definition}
    Q=-Q_{\alpha \mu \nu}P^{\alpha \mu \nu}.
\end{equation}

An equivalent and more illuminating form of the metric field equations \eqref{f(Q)-equations} is \cite{Lin-2021,Zhao-2022}
\begin{equation}
\label{f(Q)-equations-coordinate-basis}
 f_Q {\overset{\mbox{\scriptsize o}}{G}}_{\mu \nu} + \frac{1}{2}g_{\mu \nu} \bigl( Qf_Q - f)+ 2f_{QQ}\bigl(\partial_{\lambda}Q\bigr) P^{\lambda}_{\,\,\,\, \mu \nu}  = T_{\mu \nu}
\end{equation}
where ${\overset{\mbox{\scriptsize o}}{G}}_{\mu \nu}$ is the Einstein tensor associated with the Levi-Civita connection.
In this form, it becomes clear that if $f_{QQ}=0$ then $f(Q)= AQ+B$ and the theory is equivalent to general relativity (if $B = 0$) or general relativity with a cosmological constant (if $B\neq 0$). Furthermore, if the nonmetricity scalar $Q$ is constant the metric field equations \eqref{f(Q)-equations-coordinate-basis} are also equivalent to those of general relativity with a cosmological constant, but in this case for any function $f(Q)$.

Solutions to  equations \eqref{f(Q)-equations-coordinate-basis} have been investigated as well as their implications to cosmology \cite{cosmology1,cosmology2,cosmology3}. In particular, it has been found  that $f(Q)$ gravity can give rise to inflationary solutions that seem to be consistent with the current understanding of the physics of the early universe \cite{cosmology2}.

\section{G\"odel-type metrics}\label{Godeltype}

The famous rotating universe model discovered by G\"odel \cite{Godel} is the best
known example of a solution to  Einstein’s equations
with a physically well-motivated source that explicitly shows 
that general relativity allows the existence of closed timelike worldlines, implying violation of causality in certain regions of spacetime. G\"odel's model
is a solution to the Einstein field equations with a cosmological constant and  dust, but it can also be seen as a solution without cosmological constant with the matter content described by a  perfect fluid with  equation of state $p=\rho$. Because of its unforeseen properties,   G\"odel’s solution has motivated a large amount of research into 
rotating G\"odel-type models as well as on causal anomalies
both in general relativity \cite{Stockum,Som,Morris-Thorne,Morris-Thorne-AJP,Tipler,Gott,Alcubierre,Reboucas-Features,Reboucas-Time-Travel,Krasinski,Carneiro,Obukhov,Dabrowski}
and modified gravity theories \cite{Accioly,Barrow-Dabrowski,Reboucas-KK,Boyda,Barrow-Brane,Banados,Huang,Furtado,Tao-Xun,Liu-Yu,Porfirio}.

G\"odel's solution to the general relativity field equations is a particular member of the broad family of  geometries whose general form in cylindrical coordinates $(r, \phi, z)$ is \cite{Reb_Tiomno}
\begin{equation}  
\label{Godel-metric}
ds^2 = [dt + H(r)d\phi]^2 - D^2(r)d\phi^2 - dr^2 - dz^2.
\end{equation}
The necessary and sufficient conditions for this metric
to be spacetime homogeneous are \cite{Reb_Tiomno,RebAman}
\begin{equation} 
\label{ST-hom-cond}
\frac{H'}{D}  =  2\omega , \qquad 
\frac{D''}{D}  =  m^{2},
\end{equation}
where the prime denotes derivative, and the parameters $\omega,m$
are constants such that  $\omega$ is real and nonzero while $-\infty < m^{2} < \infty$.  As a consequence, the G\"odel-type metrics fall into three classes: (i) hyperbolic class ($m^2>0$); (ii) trigonometric class ($m^2 < 0$); (iii) linear class ($m^2=0$).

With  the coordinate numbering  $x^0=t,x^1=r,\, x^2=\phi,\, x^3=z$
the components of the metric tensor are
\begin{equation}
\label{metric-tensor}
\bigl(g_{\mu \nu}\bigr) = \left(\begin{array}{cccc}
               1 & 0 & H & 0\\
               0 & -1 & 0 & 0 \\
               H & 0 & G & 0\\
               0 & 0 & 0 & -1
               \end{array}
							\right)
\end{equation}
while those of  its inverse are
\begin{equation}
\label{metric-tensor-inverse}
\bigl(g^{\mu \nu}\bigr) = \left(\begin{array}{crcr}
               -G/D^2 & 0 & H/D^2 & 0\\
               0 & -1 & 0 & 0 \\
               H/D^2 & 0 & -1/D^2 & 0\\
               0 & 0 & 0 & -1
               \end{array}
							\right)
\end{equation}
where 
\begin{equation} 
\label{relation-DHG}
G = H^2-D^2.
\end{equation}
The metric \eqref{Godel-metric} can be put in the form \cite{Reboucas-tese}
\begin{equation}
\label{metric-tetrad}
ds^2 = \eta_{AB} \theta^A \theta^B = \bigl(\theta^0\bigr)^2 - \bigl(\theta^1\bigr)^2 - \bigl(\theta^2\bigr)^2 - \bigl(\theta^3\bigr)^2
\end{equation}
where $\eta_{AB} = \mbox{diag} \, (1,-1,-1,-1)$
and $\theta^A = e^A_{\mu}dx^{\mu}$. The nonvanishing components of the  tetrad $e^A_{\mu}$ are
\begin{equation}
\label{tetrad-nonzero}
 e^{(0)}_0 =1, \quad e^{(0)}_2 =H, \quad e^{(1)}_1 = 1, \quad e^{(2)}_2 = D,  \quad e^{(3)}_3 = 1.
\end{equation}
The inverse tetrad $e^{\mu}_B$, defined by $e^A_{\mu}e^{\mu}_B = \delta^A_B$, has the following nonzero components:
\begin{equation}
\label{tetrad-inverse-nonzero}
 e^0_{(0)} =e^1_{(1)} = e^3_{(3)}  = 1, \quad e^0_{(2)} =- \frac{H}{D}, \quad e^2_{(2)} = \frac{1}{D}.
\end{equation}

We shall employ these tetrads for the sole purpose of projecting the metric field equations \eqref{f(Q)-equations-coordinate-basis}, and  for the  very practical reason that the components $\,{\overset{\mbox{\scriptsize o}}{G}}_{AB}\,$ of the Einstein tensor for  G\"odel-type metrics have been previously computed  and are easy to find in the literature.  

\section{G\"odel-type metrics in $f(Q)$ gravity}\label{Godeltype-f(Q)}

Our main interest lies in finding out whether $f(Q)$ gravity admits G\"odel-like violations of causality. To this end,  
we must investigate whether there are G\"odel-type solutions in $f(Q)$ gravity. We  look for solutions to the field equations in the coincident gauge.

The coincident gauge is defined by
\begin{equation}
\label{coincident-gauge}
\Gamma^{\lambda}_{\,\,\mu \nu} =0, 
\end{equation}
implying that
\begin{equation}
\label{covariant-derivative-coincident-gauge}
\nabla_{\mu} = \partial_{\mu}. 
\end{equation}
In this gauge the nonmetricity components are simply
\begin{equation}
\label{nonmetricity-coincident-gauge}
Q_{\alpha\mu \nu} = \partial_{\alpha}g_{\mu \nu},
\end{equation}
and the only nonvanishing components for the metric \eqref{Godel-metric} are
\begin{equation}
\label{nonmetricity-coincident-gauge-nonzero}
Q_{102} = Q_{120} = H^{\prime}(r), \quad Q_{122} = G^{\prime}(r).
\end{equation}
Taking into account Eq. \eqref{P-coordinate-basis-coincident} in the Appendix, the nonmetricity scalar \eqref{nonmetricity-scalar-definition} is immediately found to be
\begin{equation}
\label{nonmetricity-scalar-coincident}
Q = \frac{{H^{\prime}}^2}{2D^2} =2\omega^2,
\end{equation}
where Eq. \eqref{ST-hom-cond} has been used. 
The fact  that $Q$ is constant will be of the utmost consequence in what follows. As pointed out in Section \ref{f(Q)gravity}, essentially the same solutions as in general relativity will be allowed.

At this moment, we wish to call attention to a very significant point that is sometimes overlooked: the coincident gauge is legitimate only if the trivial connection $\Gamma^{\lambda}_{\,\,\mu \nu} =0$ satisfies the connection field equations   \eqref{connection-equations}.

In the  case under consideration, inasmuch as all geometric quantities depend on $x^1=r$ alone, the connection field equations \eqref{connection-equations} reduce to
\begin{equation}
\label{connection-equations-reduced}
 \frac{d^2}{dr^2}\bigl( \sqrt{-g}f_Q P^{11}_{\quad \alpha}\bigr) =0.
\end{equation}
Equation \eqref{P-coordinate-basis-coincident} in the Appendix shows that $P^{11}_{\quad \alpha}=0$, implying that  $\Gamma^{\lambda}_{\,\,\mu \nu} =0$ satisfies the connection field equations.

 It is of paramount importance to emphasize that here the coincident gauge is not merely a convenient but questionable choice because the  connection field equations are satisfied. 
 This means that our forthcoming results are not an artifact of an improper gauge choice.

In the tetrad basis defined by  \eqref{tetrad-nonzero} and \eqref{tetrad-inverse-nonzero}  the metric field equations \eqref{f(Q)-equations-coordinate-basis} take the form
\begin{equation}
\label{f(Q)-equations-tetrad-basis}
 f_Q {\overset{\mbox{\scriptsize o}}{G}}_{AB} + \frac{1}{2}\eta_{AB} \bigl( Qf_Q - f)+ 2f_{QQ}U_C P^C_{\,\,\,\, AB}  = T_{AB},
\end{equation}
where 
\begin{equation}
\label{quantities-tetrad-basis}
{\overset{\mbox{\scriptsize o}}{G}}_{AB}= e^{\mu}_A e^{\nu}_B {\overset{\mbox{\scriptsize o}}{G}}_{\mu\nu}, \quad 
 U_C=e^{\lambda}_C\partial_{\lambda}Q, \quad
P^C_{\,\,\,\, AB}= e^C_{\lambda} e^{\mu}_A e^{\nu}_B P^{\lambda}_{\,\,\,\, \mu \nu}, \quad
T_{AB}= e^{\mu}_A e^{\nu}_B T_{\mu\nu}.
\end{equation}
Tetrad indices are raised with $\eta^{AB}$ and lowered with $\eta_{AB}$. We choose to write the metric field equations \eqref{f(Q)-equations-coordinate-basis} in the tetrad basis for a purely pragmatic reason: the tetrad basis components $\, {\overset{\mbox{\scriptsize o}}{G}}_{AB}\,$ of the Einstein tensor for G\"odel-type metrics can be readily found in the literature, sparing us from unnecessary tiresome calculations. Furthermore, the form taken by the Einstein tensor in the tetrad basis is extremely simple: $\, {\overset{\mbox{\scriptsize o}}{G}}_{AB}\,$ is diagonal and all of its components are constants --- see Eq. \eqref{Einstein-tensor-tetrad} below.

Since $Q$ is constant,  $U_C=e^{\lambda}_C\partial_{\lambda}Q =0$ and the field equations \eqref{f(Q)-equations-tetrad-basis} reduce to
\begin{equation}
\label{f(Q)-equations-tetrad-basis-reduced}
 f_Q {\overset{\mbox{\scriptsize o}}{G}}_{AB} + \frac{1}{2}\eta_{AB} \bigl( Qf_Q - f)  = T_{AB}.
\end{equation}
For the G\"odel-type geometry \eqref{Godel-metric} with  \eqref{ST-hom-cond} and \eqref{relation-DHG},  the nonzero components of the Einstein tensor in the tetrad basis are \cite{Reboucas-tese,Santos-Reb}
\begin{equation}
\label{Einstein-tensor-tetrad}
{\overset{\mbox{\scriptsize o}}{G}}_{(0)(0)} = 3\omega^2 -m^2,  \quad {\overset{\mbox{\scriptsize o}}{G}}_{(1)(1)} =  {\overset{\mbox{\scriptsize o}}{G}}_{(2)(2)} = \omega^2, \quad 
 {\overset{\mbox{\scriptsize o}}{G}}_{(3)(3)} = m^2 - \omega^2.
\end{equation}

\section{Perfect fluid and violation of causality}\label{Perfectfluid}

The G\"odel-type metric \eqref{Godel-metric} is compatible with the perfect fluid energy-momentum tensor
\begin{equation}
\label{ energy-momentum-tensor}
T_{AB} = (\rho + p) u_Au_B - p \eta_{AB}, \qquad u_A = (1,0,0,0),
\end{equation}
whose nonzero components are 
\begin{equation}
\label{ energy-momentum-tensor-nonzero-components}
T_{(0)(0)} = \rho, \qquad T_{(1)(1)} = T_{(2)(2)}  =T_{(3)(3)}  = p.
\end{equation} 
In this case, the nontrivial components of the field equations \eqref{f(Q)-equations-tetrad-basis-reduced} for $f(Q)$ gravity take the following form:
\begin{eqnarray}
\label{field-equations-tetrad-coincident-gauge00}
 f_Q(3\omega^2 - m^2) +\frac{1}{2}( Qf_Q-f)  & = & \rho,\\
 \label{field-equations-tetrad-coincident-gauge11}
 f_Q\omega^2 - \frac{1}{2} ( Qf_Q-f)  & = & p,\\
 \label{field-equations-tetrad-coincident-gauge22}
 f_Q\omega^2 -\frac{1}{2}\bigl( Qf_Q-f) & = & p,\\
 \label{field-equations-tetrad-coincident-gauge33}
 f_Q(m^2 -\omega^2) -\frac{1}{2} ( Qf_Q-f)  & = & p.
\end{eqnarray}

By combining Eqs. \eqref{field-equations-tetrad-coincident-gauge11} and \eqref{field-equations-tetrad-coincident-gauge33} we get
\begin{equation}
\label{fQ-m2=2omega2}
f_Q\omega^2=f_Q(m^2-\omega^2)
\end{equation}
which implies
\begin{equation}
\label{m2=2omega2}
 m^2 = 2\omega^2
\end{equation}
on the assumption that $f_Q > 0$. The equality $m^2=2\omega^2$, in turn,
implies that the metric \eqref{Godel-metric} reduces to G\"odel's original solution \cite{Godel,Reb_Tiomno}.
In general, if  $0 < m^2 < 4\omega^2$ there is a  radius beyond which causality may be violated \cite{Godel,Reboucas-tese,Reb_Tiomno}. This critical radius is
\begin{equation}
\label{critical-radius-general}
r_c= \frac{2}{\vert m \vert} \sinh^{-1}\left( \frac{4\omega^2}{m^2}-1\right)^{-1}.
\end{equation}
Therefore, taking \eqref{m2=2omega2} into account,   it follows  that in the present case the critical radius  is 
\begin{equation}
\label{critical-radius}
r_c= \frac{2}{\sqrt{2}\,\vert \omega \vert} \, \sinh^{-1}(1) = \sqrt{2}\ln (1+\sqrt{2})\,\vert \omega \vert^{-1}.
\end{equation}
There is violation of causality for $r>r_c$  because the circles 
\begin{equation}
r=\mbox{const}>r_c,\qquad  t =\mbox{const}, \qquad z=\mbox{const}
\end{equation}
are closed timelike curves \cite{Godel,Reboucas-tese,Reb_Tiomno}.

Note that if $Q=Q_0$ and $f_Q(Q_0)=0$ then Eq. \eqref{fQ-m2=2omega2} does not fix a relation between $m$ and $\omega$. This would be the case, for example, if $f(Q)=Q_0 + a(Q-Q_0)^2$ with nonzero constants $Q_0$ and $a$.   Then Eqs. \eqref{field-equations-tetrad-coincident-gauge00} to \eqref{field-equations-tetrad-coincident-gauge33} would hold true for a fluid with equation of state $p=-\rho$, which is tantamount to  a cosmological constant, for any values of $m$ and $\omega$. Of course, this still includes all cases such that $0<m^2<4\omega^2$, which entail violation of causality.

\section{Energy conditions}\label{Energycond}

Energy conditions are essentially  statements to the effect that the total energy density is nonnegative everywhere in spacetime and that matter cannot travel faster than light \cite{Hawking,Carroll}.  They are expected to be obeyed by ``normal'' matter and non-gravitational fields in order to rule out  ``unphysical'' solutions of the Einstein field equations.

Let us examine to what extent the function $f(Q)$ is restricted by  energy conditions on the perfect fluid that acts as the matter source of G\"odel's solution in the context of $f(Q)$ gravity. 

The weak energy condition requires $\rho \geq 0$ and $\rho + p \geq 0$. From Eq. \eqref{field-equations-tetrad-coincident-gauge00} we must have 
\begin{equation}
\label{energy-condition1}
 2Qf_Q-f \geq 0,  
\end{equation}
where Eqs. \eqref{nonmetricity-scalar-coincident} and \eqref{m2=2omega2} have been used.  In addition to that, with the use of Eqs. \eqref{field-equations-tetrad-coincident-gauge00} and 
\eqref{field-equations-tetrad-coincident-gauge11} the condition $\rho + p \geq 0$ implies
\begin{equation}
\label{energy-condition2}
 Qf_Q \geq 0.
\end{equation}
Together with $\rho + p \geq 0$, the strong  energy condition stipulates that  $\rho + 3p \geq 0$ . From \eqref{field-equations-tetrad-coincident-gauge00} and 
\eqref{field-equations-tetrad-coincident-gauge11} it follows that the latter condition leads to 
\begin{equation}
\label{energy-condition3}
 Qf_Q + f \geq 0.
\end{equation}
Since $Q>0$, condition \eqref{energy-condition2} is satisfied if $f_Q>0$. Therefore, the weak and strong energy conditions are certainly obeyed as long as  the function $f(Q)$ satisfies the following inequalities:
\begin{equation}
\label{energy-conditions-f(Q)}
 f_Q>0, \qquad 2Qf_Q-f \geq 0, \qquad Qf_Q + f \geq 0.
\end{equation} 
The dominant energy condition stipulates that  $\rho - \vert p \vert \geq 0$. Taking into account that $\omega^2=Q/2$ and $m^2=2\omega^2=Q$, equations \eqref {field-equations-tetrad-coincident-gauge00} and \eqref {field-equations-tetrad-coincident-gauge11} give 
\begin{equation}
\label{energy-conditions-dominant1}
 \rho = Qf_Q - \frac{f}{2}, \qquad p = \frac{f}{2}.
\end{equation} 
Therefore,
\begin{equation}
\label{energy-conditions-dominant2}
 \rho - \vert p \vert = Qf_Q - \frac{f}{2} - \frac{\vert f \vert }{2}.
\end{equation} 
If $ f \geq 0$ this reduces to $\rho - \vert p \vert =  Qf_Q - f$ whereas if $ f < 0$ one has $\rho - \vert p \vert = Qf_Q$. The latter  inequality is satisfied if $f_Q>0$, while the former is a new restriction on $f(Q)$ in addition to  \eqref{energy-conditions-f(Q)}. But notice that $2Qf_Q-f \geq 0$ is automatically satisfied if  $Qf_Q-f \geq 0$ because $Q>0$ and $f_Q>0$. 
So, all energy conditions are fulfilled if the following set of inequalities are satisfied:
\begin{equation}
\label{energy-conditions-complete-f(Q)}
 f_Q>0, \qquad Qf_Q-f \geq 0,  \qquad Qf_Q + f \geq 0.
\end{equation} 

By way of illustration, let us now see what these inequalities have to say about three models proposed in the literature.

$\bullet$ Model 1: $f(Q) = Q+ aQ^2$ (considered in \cite{Lin-2021} and also in \cite{Lavinia-BlackHole}). Conditions \eqref{energy-conditions-complete-f(Q)} require
\begin{equation}
\label{energy-conditions-f(Q)-model1}
 1+2aQ>0, \qquad aQ^2 \geq 0, \qquad 2Q + 3aQ^2 \geq 0.
\end{equation} 
All conditions are satisfied as long as $a \geq 0$.

$\bullet$ Model 2: $f(Q) = Q^n$ (considered in \cite{Lavinia-BlackHole}). For this model conditions \eqref{energy-conditions-complete-f(Q)} give
\begin{equation}
\label{energy-conditions-f(Q)-model2}
 nQ^{n-1}>0, \qquad  (n-1)Q^n \geq 0, \qquad (n+1)Q^n \geq 0.
\end{equation} 
All conditions are satisfied if $n \geq 1$.

$\bullet$ Model 3: $f(Q) = Qe^{a/Q}$ (considered in  \cite{cosmology2}). In this case conditions \eqref{energy-conditions-complete-f(Q)} yield
\begin{equation}
\label{energy-conditions-f(Q)-model3}
\bigg(1-\frac{a}{Q}\bigg) e^{a/Q} >0, \qquad   - a e^{a/Q} \geq 0, \qquad (5Q - 3a) e^{a/Q} \geq 0.
\end{equation} 
All conditions are satisfied if $a \leq 0$.

\section{Causal solutions with massless scalar field}\label{Scalar field}

A source other than a perfect fluid may lead to causal solutions. Following Rebou\c cas and Tiomno \cite{Reb_Tiomno}, let us add a massless scalar field $\Phi$ to the perfect fluid to obtain  the energy momentum tensor
\begin{equation}
\label{TAB-scalarfield}
T_{AB} = (\rho + p) u_Au_B - p \eta_{AB} + \partial_A\Phi \partial_B\Phi - \frac{1}{2}\eta_{AB}\eta^{CD}\partial_C\Phi \partial_D\Phi .
\end{equation}

\subsection{Scalar field depending  on $r$ alone}

As our first try, let us suppose $\Phi$ depends only on $r$. In this case $T_{AB}$ is diagonal and its components are
\begin{equation}
\label{TAB-scalarfield-phi-r}
T_{(0)(0)} = \rho + \frac{\Phi^{\prime}(r)^2}{2}, \quad T_{(1)(1)} = p   + \frac{\Phi^{\prime}(r)^2}{2}, \quad  T_{(2)(2)} =  T_{(3)(3)} = p   - \frac{\Phi^{\prime}(r)^2}{2}.
\end{equation}
Then the field equations \eqref{f(Q)-equations-tetrad-basis-reduced} become
\begin{eqnarray}
\label{field-equations-tetrad-coincident-gauge-scalar-r00}
 f_Q(3\omega^2 - m^2) +\frac{1}{2}( Qf_Q-f)  & = & \rho + \frac{\Phi^{\prime}(r)^2}{2},\\
 \label{field-equations-tetrad-coincident-gaugescalar-r11}
 f_Q\omega^2 - \frac{1}{2} ( Qf_Q-f)  & = & p + \frac{\Phi^{\prime}(r)^2}{2},\\
 \label{field-equations-tetrad-coincident-gaugescalar-r22}
 f_Q\omega^2 -\frac{1}{2}\bigl( Qf_Q-f) & = & p - \frac{\Phi^{\prime}(r)^2}{2},\\
 \label{field-equations-tetrad-coincident-gaugescalar-r33}
 f_Q(m^2 -\omega^2) -\frac{1}{2} ( Qf_Q-f)  & = & p - \frac{\Phi^{\prime}(r)^2}{2}.
\end{eqnarray}
From Eqs. \eqref{field-equations-tetrad-coincident-gaugescalar-r22} and \eqref{field-equations-tetrad-coincident-gaugescalar-r33} it follows at once that $m^2=2\omega^2$ and we are back to G\"odel's original model. Moreover, Eqs. \eqref{field-equations-tetrad-coincident-gaugescalar-r11} and \eqref{field-equations-tetrad-coincident-gaugescalar-r22} imply that $\Phi^{\prime}(r)=0$, and one returns to the pure perfect fluid case of Section \ref{Perfectfluid}.

\subsection{Cosmological constant and scalar field depending  on $z$ alone}

Let us take 
\begin{equation}
\label{cosmological-constant}
f(Q) = g(Q) - 2\Lambda,
\end{equation}
which is equivalent to adding a cosmological constant $\Lambda$ to the field equations, and assume not only that there is no fluid but also that $\Phi$ depends only on $z$.
In this case, putting $\rho = p =0$, the total energy momentum tensor \eqref{TAB-scalarfield} reduces to the one associated with the massless scalar field, which is  diagonal with components
\begin{equation}
\label{TAB-scalarfield-phi-z}
T_{(0)(0)} = - T_{(1)(1)} = - T_{(2)(2)} =  T_{(3)(3)} =  \frac{\Phi^{\prime}(z)^2}{2}.
\end{equation}

Then the field equations \eqref{f(Q)-equations-tetrad-basis-reduced} become
\begin{eqnarray}
\label{field-equations-tetrad-coincident-gauge-scalar-z00}
 g_Q(3\omega^2 - m^2) +\Lambda +\frac{1}{2}( Qg_Q-g)  & = & \frac{\Phi^{\prime}(z)^2}{2},\\
 \label{field-equations-tetrad-coincident-gaugescalar-z11}
 g_Q\omega^2 - \Lambda - \frac{1}{2} ( Qg_Q-g)  & = & - \frac{\Phi^{\prime}(z)^2}{2},\\
 \label{field-equations-tetrad-coincident-gaugescalar-z22}
 g_Q\omega^2 - \Lambda - \frac{1}{2}\bigl( Qg_Q-g) & = & - \frac{\Phi^{\prime}(z)^2}{2},\\
 \label{field-equations-tetrad-coincident-gaugescalar-z33}
 g_Q(m^2 -\omega^2) -\Lambda - \frac{1}{2} ( Qg_Q-g)  & = &  \frac{\Phi^{\prime}(z)^2}{2}.
\end{eqnarray}

By adding Eqs. \eqref{field-equations-tetrad-coincident-gauge-scalar-z00} and \eqref{field-equations-tetrad-coincident-gaugescalar-z11} one arrives at
\begin{equation}
\label{gQ-m2=4omega2}
g_Q (4\omega^2- m^2) =0.
\end{equation}
If $g_Q>0$ this leads to
\begin{equation}
\label{m2=4omega2}
m^2 = 4\omega^2.
\end{equation}

According to Eq. \eqref{critical-radius-general},   $r_c = \infty$ if $m^2=4\omega^2$. In other words, there is no critical radius  and the solution is causal.  With    $m^2=4\omega^2$ the field equations \eqref{field-equations-tetrad-coincident-gaugescalar-z22} and \eqref{field-equations-tetrad-coincident-gaugescalar-z33} become
\begin{eqnarray}
 \label{field-equations-tetrad-coincident-gauge-scalar-z22-m2=4omega2}
 \omega^2 g_Q - \Lambda - \frac{1}{2} ( Qg_Q-g)  & = & - \frac{\Phi^{\prime}(z)^2}{2},\\
 \label{field-equations-tetrad-coincident-gauge-scalar-z33-m2=4omega2}
3\omega^2 g_Q -\Lambda - \frac{1}{2} ( Qg_Q-g)  & = &  \frac{\Phi^{\prime}(z)^2}{2}.
\end{eqnarray}
Upon adding Eqs.  \eqref{field-equations-tetrad-coincident-gauge-scalar-z22-m2=4omega2} and \eqref{field-equations-tetrad-coincident-gauge-scalar-z33-m2=4omega2}  one gets
\begin{equation}
\label{cosmological-constant-value}
\Lambda= \frac{1}{2}\bigl( Qg_Q+ g\bigr),
\end{equation}
where we have used $\omega^2 = Q/2$.
Insertion of the above result into any of equations  \eqref{field-equations-tetrad-coincident-gauge-scalar-z22-m2=4omega2} or \eqref{field-equations-tetrad-coincident-gauge-scalar-z33-m2=4omega2} leads to
\begin{equation}
\label{cosmological-constant-value-final}
\Phi^{\prime}(z)^2 =  Qg_Q.
\end{equation}
Since $Q>0$ and $g_Q=f_Q>0$ by conditions \eqref{energy-conditions-complete-f(Q)}, the right-hand side of this equation is positive. Thus,  Eq. \eqref{cosmological-constant-value-final}  is satisfied by a scalar field of the form $\Phi (z) = az$ where $a$ is a real nonzero constant. Assuming that the function $g(Q)$ itself obeys conditions \eqref{energy-conditions-complete-f(Q)},    the cosmological constant \eqref{cosmological-constant-value} is positive. So, for any reasonable function $f(Q)$,  we have constructed a causal solution to the field equations with a
positive cosmological constant and a massless scalar field as matter source.

\section{Conclusion}\label{Conclusions}

The content of any new gravity theory should be explored to the fullest extent and depth possible. One relevant question to be asked is whether, as general relativity, the new theory admits G\"odel-like noncausal solutions.  Aiming to investigate the possible existence of G\"odel-type solutions in $f(Q)$ gravity, we have studied the  conditions for a class of homogeneous G\"odel-type metrics to satisfy  the field equations of the theory.  We have shown that adoption of the coincident gauge  is warranted because  the connection field equations are obeyed. Working in the coincident gauge we have found that    $f(Q)$ gravity admits G\"odel-type solutions for any function $f(Q)$. Out of all G\"odel-type metrics only G\"odel's original model is allowed in $f(Q)$ gravity with a perfect fluid as source. This implies that violation of causality in the form of closed timelike curves is a feature common to general relativity and  $f(Q)$ gravity.  We have also found that mild restrictions on the function $f(Q)$ arise from the imposition of energy conditions on the fluid. By considering three forms for $f(Q)$ that have been put forward in the literature, we have established that they satisfy the weak,   strong and dominant energy conditions for a wide range of values of the parameters that enter $f(Q)$.

In the presence of a cosmological constant, with a massless scalar field as matter content, we have constructed    causal solutions for any sensible function $f(Q)$. Just like general relativity, $f(Q)$ gravity admits causal or noncausal solutions depending on the matter content described by the energy momentum tensor, and on the presence or absence of a cosmological constant. 
Future work should further explore this issue, examining whether the noncausal behavior of this modified gravity theory extends to a wider class of solutions under more general physical conditions.

\begin{appendix}

\section{Computation of relevant geometric quantities}

By means of \eqref{metric-tensor-inverse}  and \eqref{nonmetricity-coincident-gauge-nonzero} the nonvanishing contravariant components of $Q^{\alpha\mu \nu}$ are found to be
\begin{eqnarray}
\label{nonmetricity-coincident-gauge-contravariant-nonzero}
Q^{100} & = & \frac{2GHH^{\prime}-H^2G^{\prime}}{D^4},\\
Q^{102} = Q^{120} & = & \frac{HG^{\prime}-(G+H^2)H^{\prime}}{D^4},\\
Q^{122}  & = & \frac{2HH^{\prime}-G^{\prime}}{D^4}.
\end{eqnarray}
From  \eqref{Q-tildeQ} it follows that  ${\tilde Q}^{\alpha} = 0$ and the only nonzero component of $Q^{\alpha}$ is
\begin{equation}
\label{Q-coordinate-basis}
 Q^{1} = \frac{G^{\prime}-2HH^{\prime}}{D^2}.
\end{equation}
Finally, equation \eqref{P-alpha-mu-nu} implies that the coordinate basis nonzero components of $P^{1 \mu\nu}$ are
\begin{equation}
\label{P-coordinate-basis-coincident}
P^{100} = \frac{G^{\prime}}{4D^2}, \qquad P^{102} = P^{120}= -\frac{H^{\prime}}{4D^2}, \qquad P^{133} = \frac{2HH^{\prime}-G^{\prime}}{4D^2}.
\end{equation}

\end{appendix}



\end{document}